\documentstyle[12pt]{article}
\makeatletter
\@addtoreset{equation}{section}

\makeatother

\newcommand{\be}{\begin{equation}}
\newcommand{\ee}{\end{equation}}

\newcommand{\reff}[1]{(\ref{#1})}

\begin{document}

\title{The quark-composites approach to QCD: The nucleon-pion system}  \author{
  { Fabrizio Palumbo~\thanks{This work has been partially 
  supported by EEC under TMR contract ERB FMRX-CT96-0045}}             \\[-0.2cm]
  {\small\it INFN -- Laboratori Nazionali di Frascati}  \\[-0.2cm]
  {\small\it P.~O.~Box 13, I-00044 Frascati, ITALIA}          \\[-0.2cm]
  {\small Internet: {\tt palumbof@lnf.infn.it}}     
   }
\maketitle

\thispagestyle{empty}   

\begin{abstract}
We present a new perturbative approach to QCD based on the use of quark composites as fundamental variables.
The composites with the quantum numbers of the  nucleons are assumed as new integration variables in the
Berezin integral which defines the partition function, while the composites with the quantum numbers of the
chiral mesons are replaced by auxiliary bosonic fields. The action is modified by the addition of irrelevant 
operators which provide the kinetic terms for these composites, and the quark action is treated as a
perturbation. The resulting expansion has the quark confinement built in. As a first application we
investigate the  pion-nucleon interaction.
 \end{abstract}

\clearpage
 
\section{Introduction}

We can divide the calculations in QCD in two classes: Those which aim at clarifying the confining mechanism, 
and those devoted to the description of hadronic physics. One of the present the difficulties in this
field stems from the fact that the former calculations are done in a framework where it would be very awkward,
if not impossible, to perform the latters. More generally, the actual understanding of the theory requires a
nonperturbative approach for low energy  and perturbative methods for high energy, which until now it has been
impossible to unify in a unique scheme.

The strategy we adopt to overcome the impasse is to use quark composites with hadronic quantum numbers as
fundamental variables~\cite{Palu93}. An earlier attempt in this spirit can be found in~\cite{Kawa}, but it is
restricted to the strong coupling. We avoid such limitation by means of new technical tools.

There is a fundamental difference between trilinear and bilinear composites. Surprisingly enough  some trilinear
composites, in particular the nucleon fields, can easily be assumed as integration variables in the Berezin
integral which defines the  partition function, since they satisfy the same integration rule as the quark
fields. As a consequence the free action of these composites is the Dirac action~\cite{DeFr}.

Also the mesonic composites can be introduced as integration variables, but the resulting integral does not
reduce in general to a Berezin or to an ordinary integral, and it is very complicated. In particular the
propagator {\it is not the inverse of the wave operator}. It has then required  some effort to find out an
 operator containing the free action of the chiral composites~\cite{Cara}, but once it has been constructed, we
can circumvent the difficulty of the integral over the chiral composites by replacing them by auxiliary fields
by means of the  Stratonovich-Hubbard representation~\cite{Hub}.

Since the free composite actions are irrelevant operators, we can freely add them to the standard action,
and derive a perturbative expansion by treating the quark action as a perturbation. The expansion 
parameters are the inverse of two dimensionless constants entering the definition of the nucleon and chiral
composites, and of the number of components of the up and down quarks. This number, called $\Omega$,
is equal to 24.  

Our approach has two quite desirable features. Firstly it is compatible with a perturbative as well as
nonperturbative regime of the gluons wr to the gauge coupling constant. For this reason we adopt a
regularization on a Euclidean lattice, which is the only one suitable for both cases. This choice  seems also
natural dealing with composites. The other remarkable aspect is that the quark
confinement is built in. This is due to the fact that our perturbative expansion is of weak
coupling for the hadrons, but, as a consequence of the spontaneous breaking of
the chiral symmetry, of strong coupling for the quarks. To any finite order, the quarks can then move
only by a finite number of lattice spacings, and therefore they can never be produced in the continuum limit,
at variance with the standard perturbation theory, where the the quarks are propagating 
particles at the perturbative level, and the confinement is essentially nonperturbative.
Let us emphasize that the hopping character of the quark propagator should not generate any
confusion with a strong coupling expansion in terms of the gauge coupling: As stated above, the gauge coupling
constant can have any value required by the dynamics, while mesons and nucleons do move in the true continuum.

One can wonder whether the Wilson term to avoid spurious quark states is at all necessary in our approach: 
Since the quarks have no poles whatsoever to finite order, why should we worry about the spurious ones? In any
case we can safely assume the Wilson term for the quarks of subleading order in our expansion parameters, so
that we can ignore it in  a first order calculation.

In the previous works the nucleons and the chiral mesons were discussed separately. Here we unify their
treatment including the em interactions, so as to be able to evaluate strong corrections to em
processes. We derive a perturbative expansion in terms of the nucleon composites and the auxiliary
fields which replace the chiral composites. The numerical coefficients of this series are given in terms of 
integrals over the gluon and quark fields. As an application we investigate the nucleon-pion interaction,
getting to lowest order a contribution to be the one-pion exchange of the Yukawa model.

The paper is organized as follows. In Section 2 we define the modified partition function we will use.
In Section 3 we define the hadron composites and their actions. In Sections 4 and 5 we report for the
convenience of the reader the results, obtained in the previous works, concerning the introduction the nucleon
fields as integration variables, and of the auxiliary fields for the chiral composites. In
Section 6 we derive our perturbative expansion. In Section 7 we show that the quarks do not propagate at the
perturbative level. In Section 8 we evaluate the nucleon-pion interaction and in Section 9 we conclude our
paper.

\section{The modified partition function}

 We assume the modified partition function 

\begin{equation}
Z = \int [dU] \int [dA]\, \exp[- S_{YM} - S_M] \int [d\overline{\lambda} d\lambda]\; \exp[-S_N -S_{C} -S_Q], 
\end{equation}
where $S_{YM}$ and $S_M$ are the Yang-Mills and Maxwell actions, $S_Q$ is the action of
the quark fields and $S_N$ and $S_C$ are irrelevant operators which provide the kinetic terms for the quark
composites with the quantum numbers of the nucleons and the chiral mesons. $\lambda$ is the quark field, the
gluon field is  associated to the link variables $U_{\mu}$ and $A_{\mu}$ is the photon field. The reason why we
have introduced the photon field explicitely, is that it is associated to different link variables in
the interactions with the quarks and the composites. Finally differentials in square brackets are understood,
as usual, as the product of differentials over the sites and the intrinsic indices, and $[dU]$ is the Haar
measure.

All the elementary fields live in an anysotropic euclidean lattice of spacing $a$ in the spatial directions and
$a_t$ in the time direction, whose sites are identified by fourvectors $x$ of spatial components $x_k=0,...N$
and time component $t=0,...2N_t$, and satisfy periodic boundary conditions, with the exception of the quark
fields which are antiperiodic in time: $\lambda(x)= \lambda(x+Ne_k)= - \lambda(x+ 2 N_t e_t)$, 
$e_k$ and $e_t$ being the unit vectors in the $k$ and time directions. 

The simplest way of unifying the treatment of nucleons and mesons is to put them at different, complementary
sites of the latttice. A possible option, which respects the hypercubic symmetry, is to put them on the
sites of two dual sublattices, defining the parallel transport along the major diagonals~\cite{Cara1}.
Here we make a different choice, possibly more suitable for the hadron termodynamics. We restrict the
nucleons/mesons to the odd/even sites in one direction ( which is natural to assume 
as the time direction), while letting them to occupy the whole remaining three dimensional volume. The
hypercubic symmetry is lost, but it remains valid at least for the free composites, if we assume the temporal
spacing half the spatial one. We will keep however the spacings unrelated having the hadron thermodynamics
in mind. So the nucleon and chiral composites are defined respectively at the sites $n=(x_k,2t+1)$ and
$c=(x_k,2t)$, $t=0,...N_t$, which identify the nucleon and chiral  lattices. Accordingly we introduce specific
notations for the  sums over the whole lattice or its sublattices  \begin{eqnarray} (f,g) & = &  a^3 a_t \sum_x
f(x) g(x), \nonumber\\
 (f,g)_n &= & a^3 2 a_t \sum_{n} f(n) g(n),\;\;\; (f,g)_c =  a^3 2 a_t \sum_{c} f(c) g(c). 
\end{eqnarray}

The quark fields $\lambda^a_{\tau \alpha}$ with color, isospin and Dirac indices $a$,
$\tau$ and $\alpha$ respectively, are related to the up and down quarks
according to
\be
\lambda^a_{1\alpha}=u^a_{\alpha},\;\;\;\lambda^a_{2\alpha}=d^a_{\alpha}.
\ee
Their action is
\be
S_Q= (\overline{\lambda}, Q \lambda ),
\ee
where
\begin{eqnarray}
Q_{x,y} & = &  M_Q 
 \delta_{xy} -{1\over 2a}\sum_k  
  (r_q-\gamma_k) U_k(x)v_k(x) \delta_{x+e_k,y}
\nonumber\\
 & & -{1\over 2a_t} \sum_{\epsilon} 
  (r_q-\gamma_{\epsilon t}) U_{\epsilon t}(x) v_{ \epsilon t}(x) \delta_{x + e_{\epsilon t},y},
\;\;\;\epsilon=\pm 1, 
\end{eqnarray}
with
\be
M_Q= m_q + r_q \left({3 \over a} +{1 \over a_t} \right) .
\ee
We have adopted the standard conventions
\begin{eqnarray}
 & & k  \in  \{ -3,\ldots,3\},\;\;\;e_{-k}=-e_k,\;\;\;e_{-t} = - e_t, 
\nonumber\\
 & & \gamma_{-k}  =  - \gamma_k,\;\;\;\gamma_{-t} = - \gamma_t, 
\nonumber\\
 & & U_{-k}(x)  =  U^+_k(x-e_k),\;\;\;U_{-t}(x)  =  U^+_t(x-e_t),  \label{conv}
\end{eqnarray}
and the corresponding ones for the link variables $v_{\mu}$ associated to the photon field
\be
v_{k \tau}(x) = \exp[i a q_{\tau} A_k(x)],\;\;\;v_{t \tau}(x) = \exp[i a_t q_{\tau}A_t(x)].
\ee
 $q_{\tau}$ is the quark electric charge
\be
q_1={ 2\over 3} e,\;\;\;q_2= -{1 \over 3} e,\;\;\;e= \mbox{electron charge}.
\ee
The spurious quark states are prevented by the Wilson term with Wilson parameter $r_q$.

Since we will assume the nucleon and chiral composites as fundamental variables, we must separate the quark
action into the terms which live only in the nucleon and chiral lattices and those which couple them
\be
S_Q = S_{Qnc}+S_{Qcn} + S_{Qn}+S_{Qc}
\ee
where
\begin{eqnarray}
S_{Qnc} &=&  a^3 a_t \sum_{n c} \overline{\lambda}(n) Q_{n,c} \lambda(c),
\nonumber\\
S_{Qcn} &=&  a^3 a_t \sum_{c n} \overline{\lambda}(c) Q_{c,n} \lambda(n),
\nonumber\\
S_{Qn} &=& { 1\over 2}(\overline{\lambda}, Q \lambda)_n,
\nonumber\\
S_{Qc} &=& { 1\over 2}(\overline{\lambda}, Q \lambda)_c.
\end{eqnarray}

We then have
\begin{eqnarray} 
   Z  & = & \int
   [dA] \exp[-S_M ] \int [d \bar{\lambda} d\lambda]_n \exp(-S_N -S_{Qn})\int [dU]\exp[-S_{YM}]
  \nonumber \\
  & & \int [d \bar{\lambda} d\lambda]_c
    \exp   \left[ -S_C - S_{Qc} - S_{Qnc} - S_{Qcn} \right],  
\end{eqnarray}
where
\be
[d\overline{\lambda} d\lambda] = [d\overline{\lambda} d\lambda]_n
[d\overline{\lambda} d\lambda]_c  \; ,
\ee

The formalism for a nonlinear change of variables in Berezin integrals has been developed in 
refs.~\cite{Palu93},~\cite{DeFr}.
In the following section we will report what is necessary to arrive at the formulation of our perturbative
expansion.

\section{The hadron composites and their actions}

 The nucleon composites are~\cite{Ioff}
\begin{equation}
\psi_{\tau\alpha}= -{2\over3}k_N^{1/2}a^3 \epsilon_{a_1a_2a_3}
\delta_{\tau\tau_2}\epsilon_{\tau_1\tau_3} (\gamma_5
\gamma_{\mu})_{\alpha\alpha_1} ({\cal C} \gamma_{\mu})_{\alpha_2\alpha_3}
\lambda^{a_1}_{\tau_1\alpha_1} 
\lambda^{a_2}_{\tau_2\alpha_2} \lambda^{a_3}_{\tau_3\alpha_3}. 
\end{equation}

In the above equation and in the sequel the summation over repeated indices is
understood, ${\cal C}$ is the charge conjugation matrix, and
 $\psi_{1\alpha},\psi_{2\alpha}$ are the proton, neutron fields.
 It is easy to check that with the above definition they transform like the quarks
under isospin, chiral and O(4) transformations.
 
We should explain why we have included in the definition the cubic power of the
lattice spacing $a$ and the parameter $k_N$. The cubic power of a parameter with the
dimension of a length, say $l^3$, is necessary to give the nucleon field the canonical
dimension of a fermion field. At the same time a power of the lattice spacing at least cubic is
necessary to make the kinetic term (with the Dirac action), irrelevant. We have written for later convenience
$l^3$ in the form $k_N^{1/2} a^3$, where $k_N$ is dimensionless and it must not diverge in the continuum limit.

 The nucleon free action is the Dirac action
\begin{equation}
S_N = ( \bar{\psi}, N \psi )_n,
\end{equation}
whose wave operator $N$ is given by
\begin{eqnarray}
N_{n_1,n_2} &=& \left[ m_N+  r_N \left( {3\over a}+ {1\over 2a_t} \right)\right] \delta_{n_1,n_2}
-{1\over 2a}\sum_k (r_N-\gamma_k)V_k(n_1)\delta_{n_1 + e_k,n_2} 
 \nonumber\\
 &  &-{1\over 4a_t}\sum_{\epsilon}(r_N-\gamma_{\epsilon t})V_{\epsilon t}(n_1)
\delta_{n_1 + 2e_{\epsilon t},n_2}.  
\end{eqnarray}
In the above equation $r_N$ is the Wilson parameter
\be
0 < r_N \leq 1, \label{restr}
\ee
$m_N$ is the mass of the nucleon and $V_{\mu}$ is the link variable associated to the e.m. field acting on
the nucleons 
\begin{eqnarray}
V_{k\tau}(n) &=& \exp ( ie_{\tau} a A_k(n) ),
\nonumber\\
V_{t\tau}(n) &=& \exp ( ie_{\tau} a_t A_t(n))\exp ( ie_{\tau} a_t A_t(n+e_t)),\;\;
e_1=e,\;\;e_2=0, 
\end{eqnarray}
with the standard conventions. Needless to say, the coupling with the em field is necessary to preserve
the abelian gauge invariance of the QCD action.

The chiral composites are the pions and the sigma
\begin{equation} 
  \pi_{\pm} = i\,k_{\pi}\,a^{2}\,\overline{\lambda} \gamma_5 { 1\over \sqrt 2}(\tau_1 \pm i \tau_2 )
\lambda,\;\;\; \pi_0 = i\,k_{\pi}\,a^{2}\,\overline{\lambda} \gamma_5 \tau_3 \lambda,\;\;\;  
\sigma = k_{\pi}\,a^{2}\,\overline{\lambda} \lambda,   
\end{equation}
where $\gamma_5$ is assumed hermitean, the $\tau_{k}$'s are the Pauli
matrices and the factor $ a^2 k_{\pi}$ has been introduced with the same criterium used
for the nucleons, so that also $ k_{\pi}$ must not diverge in the continuum limit.

The chiral transformations over the quarks
\begin{eqnarray}
\delta \lambda &=& {i\over 2} \gamma_{5} \vec{\tau}\cdot \vec{\alpha} 
\lambda \\
\delta \overline{\lambda} &=& {i\over 2} \overline{\lambda} \gamma_{5} 
\vec{\tau}\cdot \vec{\alpha} 
\end{eqnarray}
induce $O(4)$-transformations over the mesons
\begin{eqnarray}
\delta \sigma &=& \vec{\alpha} \cdot \vec{\pi} \\
\delta \vec{\pi} &=& - \vec{\alpha} \sigma.
\end{eqnarray}
For any real vectors $\pi, \chi$, we adopt the convenction
\be
\vec{\pi} \cdot \vec{\chi} = \pi_+ \chi_- + \pi_- \chi_+  +\pi_0 \chi_0.
\ee

Since for massless quarks the QCD action is chirally invariant, the action of the chiral mesons must be,
apart from a linear breaking term, $O(4)$ invariant. It must then have the form
\begin{equation}
S_{C} =  \left[ {1 \over 2} (\vec{\pi},C\vec{\pi})_c  + {1 \over 2}
(\sigma,C\sigma)_c  - { 1 \over a^2} \left(  \sqrt {\Omega} m, \sigma \right)_c \right]. \label{actionC2}  
\end{equation}
The factor $\sqrt \Omega$ has been introduced for later convenience.
The heuristic considerations which led to the choice of the wave operator $C$ are based on experience
with simple, solvable models, and can be found in~\cite{Cara}
\be
 C(w) = a^{-2}{\rho^4 \over a^2 {\cal D}^2 - \rho^2}. 
\ee
$C$ is given in terms of the covariant laplacian ${\cal D}^2$ whose action on a function $g$ is
\be
({\cal D}^2 g )(c) = {1 \over  a^2} \sum_k[ w_k(c)  g(c+e_k) - g(c)]+ 
{1 \over 4 a_t^2} \sum_{\epsilon}[ w_{\epsilon t}(c) g(c+2e_{\epsilon t}) - g(c)] 
\ee
$w$ being the link variable associated to the em field acting on the charged pions
\begin{eqnarray}
w_k (c) \pi_{\pm} &=&  \exp (\mp ie a A_k(c) )\pi_{\pm},
\nonumber\\
w_t (c) \pi_{\pm} &=&\exp (\mp ie a_t A_t(c) )\exp (\mp ie a_t A_t(c+ e_t) )\pi_{\pm}.
\end{eqnarray}
Notice that the $\pi_{\pm}$ has negative/positive charge.
A posteriori the choice of the wave operator can be justified by observing that in the Stratonovitch-Hubbard
representation the kinetic term of the auxiliary fields is related to $C^{-1}$, and therefore the present form
generates in the simplest way a Klein-Gordon operator.
Let us stress that if the $\pi$' and the $\sigma$ were ordinary bosons, rather than even Grassmann
variables, with the above action they would be static. This is not a paradox, because the propagator of
even Grassmann fields is not the inverse of the wave operator.

 The irrelevance of the nucleon action is ensured by the insertion of the appropriate
powers of the lattice spacing in the definition of the nucleon composites. For the chiral action instead the
request of irrelevance constrains also the parameter $\rho$ which cannot vanish in the continuum limit. The
dependence on $a$ assumed below for the breaking parameter $m$, ensures also the irrelevance of the chiral
symmetry breaking term.

\section{The nucleon composites as integration variables}

In this Section we report the results we will need on the change of variables for trilinear composites.

There are 8 nucleon field components at any site $n$. Let us denote by $\Psi(n)$ their product in the order
\be
\Psi(n)=\psi_{11}(n)\psi_{12}(n)...\psi_{23}(n)\psi_{24}(n).
\ee
These fields obey the same multiplication rules of the quark fields. We can then define for them an integral
\be
\int \prod_{\tau,\alpha}d\psi_{\tau,\alpha}(n) \Psi(n) = 1, \;\;\;\ \mbox{all the other integrals vanishing},
\ee
which obeys the same rule as the Berezin integral for the quarks 
\be
\int \prod_{a,\tau,\alpha}d \lambda^a_{\tau,\alpha}(n) \Lambda(n) =1, \;\;\; \mbox{all the other integrals
vanishing}. \ee
In the latter equation $\Lambda(n)$ is the product of all the quark components at the site $n$ in the order
\be
\Lambda(n)=P(\lambda_{11}(n))...P(\lambda_{14}(n))P(\lambda_{21}(n))...P(\lambda_{24}(n)),
\ee
with
\be
P(\lambda_{\tau \alpha}(n))=\lambda^1_{\tau \alpha}(n)\lambda^2_{\tau \alpha}(n)
\lambda^3_{\tau \alpha}(n).
\ee
It is then easy to check that for an arbitrary function which depends on the quark fields only through the
$\psi$, we have, with the appropriate ordering for the differentials 
\be
\int \prod_{a,\tau,\alpha}d \lambda^a_{\tau,\alpha}(n) g(\psi(\lambda(n))) = 
J \int \prod_{\tau,\alpha}d\psi_{\tau,\alpha}(n) g(\psi(n)),
\ee
provided J, defined by
\be
\Psi(n) = J \Lambda(n),
\ee 
is different from zero. This is in fact true for the nucleons and, for the color group $SU(3)$ in 4
dimensions   
\be
J=k_N^4  \; a^{24} \; 2^{22} \cdot 3^3  \cdot 5.
\ee

This result justifies the choice of the Dirac action as the free action for the nucleons. Let us in fact define
the free nucleon partition function
\be
Z_N = \int [d \overline{\lambda} \lambda]_n \exp[-S_N].
\ee
Notice that the integral is over the quark fields in the nucleon lattice only. Let us
now consider the  nucleon-nucleon correlation functions
\be
<\overline{\psi}_{\tau \alpha}(x) \psi_{\sigma \beta}(y)> = 
{ 1\over Z_N} \int [d \overline{\lambda} \lambda]_n 
\overline{\psi}_{\tau \alpha}(x) \psi_{\sigma \beta}(y)\exp[-S_N].
\ee
Since the integrand is a function of the quark fields only through the $\psi$, we can assume them as
integration variables getting 
\be
 <\overline{\psi}_{\tau \alpha}(x) \psi_{\sigma \beta}(y)>=
- { 1\over a^3 2a_t} (N^{-1})_{\sigma \beta , \tau \alpha }(y-x),
\ee 
namely the free correlation function of two Dirac particles.
It should be noted that this result does depend neither on the value of the jacobian, provided it is different
from zero, nor on the value of $k_N$. Inverse powers of this latter constant,  however, will appear in the
perturbative expansion of QCD, due to the presence of $S_Q$ in the total partition function. 

It is remarkable, and it is at the basis of our perturbation theory, that more general formulae hold,
which involve the quark as well as the nucleon fields. In fact the following substitution rule holds in
Berezin integrals

 \begin{eqnarray}
\overline{\lambda}_{\overline{h}_1}(\overline{n})...\overline{\lambda}_{\overline{h}_l}(\overline{n})
\lambda_{h_1}(n)... \lambda_{h_{l'}}(n) &
\sim & \delta_{l,3m} \delta_{l',3m'} 
\overline{f_{\overline{I}_1...\overline{I}_m \; \overline{h}_{3m}... \overline{h}_1}} 
f_{I_1...I_{m'}\; h_1...  h_{3m'}}
\nonumber\\
& &
\overline{\psi_{\overline{I}_1}(\overline{n})...
\psi_{\overline{I}_m}(\overline{n})  } \; \psi_{I_1}(n)...
\psi_{I_{m'}}(n),   \;\label{subst} 
\end{eqnarray}
where the symbol $\sim$ does not refer to any approximation, but it means that the equality holds only under the
Berezin integral, and the $f$' are numerical coefficients called transformation functions.
We have used the shorthand notation $\overline{h}_i=(\overline{a}_i,\overline{\tau}_i,\overline{\alpha}_i),
h_i=(a_i, \tau_i, \alpha_i)$, $\overline{I}=(\overline{\tau},\overline{\alpha}), I=( \tau,\alpha)$. Restricting
ourselves to integrals trilinear in the quark and antiquark fields 
\begin{eqnarray}
 & &\int \prod_{a,\tau,\alpha}d \overline{\lambda}_{a,\tau,\alpha}(\overline{n})  d \lambda_{a,\tau,\alpha}(n)
g(\overline{\psi}(\overline{n}),\psi(n))
\overline{\lambda}_{\overline{h}_1}(\overline{n})\overline{\lambda}_{\overline{h}_2}(\overline{n})
\overline{\lambda}_{\overline{h}_3}(\overline{n})
\lambda_{h_1}(n)\lambda_{h_2}(n)\lambda_{h_3}(n) \nonumber\\
& & =   
 \overline{f_{\overline{I}\; \overline{h}_3 \overline{h}_2 \overline{h}_1}} f_{I\; h_1 h_2 h_3} 
J^2 \int \prod_{\tau,\alpha} d\overline{\psi}_{\tau,\alpha}(\overline{n}) d \psi_{\tau,\alpha}(n)
g(\overline{\psi},\psi) \overline{\psi}_{\overline{I}}(\overline{n}) \; \psi_I(n) . 
\end{eqnarray}
The transformation functions in such a case are 
\begin{equation}
f_{\tau \alpha, \tau_1 \alpha_1,\tau_2 \alpha_2,\tau_3 \alpha_3,a_1,a_2,a_3}=
\epsilon_{a_1a_2a_3}
h_{\tau\alpha,\tau_1 \alpha_1,\tau_2 \alpha_2, \tau_3 \alpha_3} , \label{ff}
\end{equation}
where
\begin{eqnarray}
h_{\tau \alpha \tau_1 \alpha_1 \tau_2 \alpha_2 \tau_3 \alpha_3}  &=& { 1\over 96}
a^{-3}k_N^{-1/2}  \left[\delta_{\tau \tau_2} \epsilon_{ \tau_1 \tau_3}
(\gamma_5 \gamma_{\mu})_{\alpha_1 \alpha} ( \gamma_{\mu} {\cal C}^{-1})_{ \alpha_2 \alpha_3}\right.
\nonumber\\
 & & \left. + \delta_{\tau \tau_1} \epsilon_{ \tau_2 \tau_3} 
 (\gamma_5 \gamma_{\mu})_{\alpha_2\alpha}
(\gamma_{\mu} {\cal C}^{-1})_{ \alpha_1 \alpha_3} \right]. \label{h}
\end{eqnarray}
The functions $h$ are totally symmetric wr to the exchange of any pair of numbered indices. The appearance 
of $ k_N^{-1/2}$ in the rhs shows how an expansion in this parameter is generated.

 It should now be clear in which sense we can talk of a change of variables. Even though the quarks cannot
be expressed in terms of the composites, we only need to invert monomials in the quark and
antiquark fields of order $3 \cdot$integer, and this can be done according to Eq.\reff{subst}.

\section{The auxiliary fields for the chiral composites}

In this Section we review the properties of the chiral action necessary for the further developments. To this
end we define the chiral partition function $Z_C$   
\be
 Z_C   =  \int [dA] \exp[-S_M]  \int [d \bar{\lambda} d\lambda]_c \exp[-S_C]
\ee
where there appear the chiral composites and the Maxwell field only. Also the chiral composites can be assumed
as integration variables, but the resulting integral is impracticable. Since $S_C$ is quadratic and the
wave operator $C$ is negative definite, the difficulty can be overcome by introducing auxiliary fields by means
of the Stratonovich-Hubbard transformation 
\begin{eqnarray} 
 \exp[-S_C ] & = & \left(\det C(A) \right)^{-1}  \int\left[{d\vec{\chi}\over\sqrt{2 \pi}}\right]_c  
   \left[{d\phi\over\sqrt{2 \pi}}\right]_c 
\nonumber\\
& & \exp \left\{ {1\over 2 } \rho^4  \left[ (\vec{\chi},(a^4 C)^{-1}\vec{\chi})_c  
+ (\phi,(a^4 C )^{-1} \phi)_c \right]  \right\}
\nonumber\\
& & \exp  \left\{ \left[{1 \over a^2}\rho^2 (\vec{\chi},\vec{\pi})_c
+ {1 \over a^2} \left(\rho^2 \phi+ \sqrt{\Omega } m,\sigma
\right)_c  \right] \right\}.
\end{eqnarray}
In the above equation we have ignored, as we will do in the sequel, field independent factors. Now the chiral
partition function 
\begin{eqnarray} 
 Z_C  & = & 
    \int [dA] \exp[-S_M - \mbox{Tr} \ln  ( C(0)^{-1}C(A) )] 
  \nonumber \\
& &  \int
   \left[{d\vec{\chi}\over\sqrt{2 \pi}}\right]_c  
   \left[{d\phi\over\sqrt{2 \pi}}\right]_c 
\exp \left\{ {1\over 2 } \rho^4 \left[ (\vec{\chi},(a^4 C)^{-1}\vec{\chi})_c  
+ (\phi,(a^4 C )^{-1} \phi)_c \right]  \right\}
\nonumber\\
& & \int [d \bar{\lambda} d\lambda]_c
 \exp  \left\{ \left[{1 \over a^2}\rho^2 (\vec{\chi},\vec{\pi})_c
+ {1 \over a^2} \left(\rho^2 \phi+ \sqrt{\Omega } m,\sigma \right)_c  \right] \right\},
\end{eqnarray}
describes bosonic fields interacting with the quarks. To get the effective action of the chiral
mesons we integrate out the quark fields with the result
\be
 Z_C   = 
   \int [dA] \exp[-S_M - \mbox{Tr} \ln ( C(0)^{-1}C(A) )]  
\int \left[{d\vec{\chi}\over\sqrt{2 \pi}}\right]_c  
   \left[{d\phi \over\sqrt{2 \pi}}\right]_c \exp[-S_{\chi}],
\ee 
where
\begin{eqnarray}
S_\chi & = & -{1\over 2 } \rho^4 \left[ (\vec{\chi},(a^4 C)^{-1}\vec{\chi})_c  
+ (\phi,(a^4 C)^{-1} \phi)_c \right] 
\nonumber\\
  & & - { \Omega \over 2} \sum_c \ln \left\{ a^2 k_{\pi}^2 \left[ \left(
    \sqrt{ \Omega} m + \rho^2 \phi(c) 
   \right)^{2} +  \rho^4 \vec{\chi}(c)^{2} \right] \right\}.  \label{schi}
\end{eqnarray}
Since $\Omega$ is a rather large number we can  apply the saddle-point method 
and evaluate the partition function as a series in inverse powers of this
parameter.
The minimum of $S_{\chi}$ is achieved for $\vec{\chi}=0$ and
\be
 \overline{\phi} ={ \sqrt{\Omega }\over \sqrt{a2a_t} \; \rho}  \left[ \sqrt{ 1+ \xi^2} - \xi\right], 
\label{barfi}
\ee
where 
\be 
\xi= { 1 \over 2 \rho} \sqrt{a 2a_t}m.
\ee
The quadratic part of $S_{\chi}$ is
\be
S_{\chi}^{(2)} = {1 \over 2}(\vec{\chi},(-{\cal D}^2 + m_-^2)\vec{\chi} )_c
+{1 \over 2}(\theta,(-{\cal D}^2 + m_+^2) \theta)_c,
\ee
where 
\be
\theta = \phi - \overline{\phi}
\ee
is the fluctuation of the field  $\phi$ and
\begin{eqnarray}
 m^2_- &=&  {2\rho^2  \over a ^2 } \; { \xi \over  b },
\nonumber\\
m^2_+ &= & {2\rho^2 \over a ^2 } \; { \sqrt{1+\xi^2} \over b },
\end{eqnarray}
with
\be
b=\sqrt{ 1+ \xi^2} + \xi.
\ee
Therefore, if we assume
\be
m = { 1 \over \rho} \sqrt{ a\over 2a_t} \; a m_{\pi}^2 + O( 1/ \sqrt \Omega ),
\ee
in the continuum limit
\begin{eqnarray}
 m^2_- &=&   \rightarrow m_{\pi}^2,
\nonumber\\
m^2_+ &= &  \rightarrow {2 \rho^2 \over a^2 },
\end{eqnarray}
and the propagator of the pion field to leading order in $1/ \sqrt{\Omega}$ turns out
to be the canonical one
\be
<\pi_h(c_1) \pi_k(c_2)> =  {1\over a^3 2a_t} \,\left({ 1 \over { -{\cal D}^2+
m_{\pi}^2}} \right)_{c_1,c_2},
\ee
while the $\sigma$ is unphysical because its mass is divergent. Notice that, in analogy to the case of
the nucleons, the constant $k_{\pi}$ is uninfluent at this stage, but its inverse powers
will appear in our  perturbative expansion of QCD.

The nucleon action $S_N$ gives an exactly free propagator. Instead, the chiral action contains a residual
interaction $ S_{\chi}^{I}$ which can be obtained by expanding the $\ln$ in inverse powers of
$\Omega$. Here we report the first terms, given in~\cite{Cara} (a factor $- { 1\over 2} \Omega$ is missing
in front of $S^{(3)}$ and $S^{(4)}$ appearing in Eq.(63) of this paper)
\begin{eqnarray}
 S_{\chi}^{I} &=&
{ 1\over 3 \sqrt{\Omega}} \left({ \rho \over b}\right)^3  { 1\over a}\sqrt{2a_t \over a}\; 
\left[ ( \theta,(-\theta^2 + 3 (\vec{\chi})^2)_c \right] 
\nonumber\\
 & & + { 1\over 4 \Omega}  \left({ \rho \over b}\right)^4 { 2a_t \over a} 
\left[(\theta^2,\theta^2)_c +( \vec{\chi}^2,(\vec{\chi}^2 -6  \; \theta^2) )_c  \right] 
 + O(\Omega^{-3/2}).
\end{eqnarray}

\section{The perturbative expansion}

In this section we derive a perturbative expansion for the partition function of QCD in terms of the nucleon
composites and the auxiliary fields. We start by the Stratonovich-Hubbard transformation to get rid of the
chiral composites 
\begin{eqnarray} 
 \exp[-S_C -S_{Qc}] & = & \left(\det C(A) \right)^{-1} \int \left[{d\vec{\chi}\over\sqrt{2 \pi}}\right]_c  
   \left[{d\phi\over\sqrt{2 \pi}}\right]_c \exp [-S_{\chi}]
\nonumber\\
& & \bigtriangleup
 \exp   \left [ {1 \over a}(\overline{\lambda},( D + Q_H )\lambda )_c \right],  \label{mass}
\end{eqnarray}
where
\begin{eqnarray}
\bigtriangleup &=& (\prod_c \det D_c)^{-1}
\nonumber\\
D &=& a  k_{\pi}\left[\rho^2\phi + \sqrt \Omega m + i \rho^2  \gamma_5 \vec{\tau} \cdot \vec{\chi}\right] 
\nonumber\\
& = & \sqrt{\Omega} \rho  b \sqrt{{a \over 2a_t}} k_{\pi} \left[ 1+ {\rho \over b}
{ \sqrt{a 2 a_t}\over \sqrt{\Omega} } \left( \theta + i \gamma_5 \tau \cdot \chi \right) \right],
\nonumber\\
(Q_H)_{c_1c_2} & = & {1\over 4 }\sum_k  (r-\gamma_k) U_k(c_1)v_k(c_1) \delta_{c_1+e_k,c_2}.
\end{eqnarray}
We have absorbed the term
 $ M_Q $ in the parameter $m$ which now reads
\be
m = { 1 \over \rho} \sqrt{a\over 2a_t} \; a m_{\pi}^2 - { 1 \over 2\sqrt \Omega k_{\pi}} M_Q.
\ee
The partition function is so transformed into 
\begin{eqnarray} 
   Z  & = & 
   \int [dA]  \exp[-S_M - \mbox{Tr} \ln ( C^{-1}(0) C(A) )]  \int [d \bar{\lambda} d\lambda]_n \exp[-S_N ]
  \nonumber \\
   & & \int
   \left[{d\vec{\chi}\over\sqrt{2 \pi}}\right]_c  
  \left[{d\theta\over\sqrt{2 \pi}}\right]_c {\cal I} \exp[-S_{\chi}],
 \end{eqnarray}
where
\begin{eqnarray} 
 {\cal I}& =& \int [dU] \exp[-S_{YM}] \exp[-S_{Qn}]\bigtriangleup \int [d \bar{\lambda} d\lambda]_c
\nonumber\\
   & & \exp   \left[{1 \over a} (\overline{\lambda},(D +Q_H)\lambda)_c  - S_{Qnc} -S_{Qcn}\right].  
\end{eqnarray}

To perform the integral over the quark fields  we expand ${\cal I}$ with respect to $
S_Q$. We note that only terms with an equal number of factors $S_{Qnc}$ and $S_{Qcn}$ can contribute.
Moreover, after the integration in the chiral lattice there remain in ${\cal I}$ only polynomials of the quark
fields in the nucleon lattice. They will contribute, according to Eq.~\reff{subst}, only if they are of order
3$\cdot$integer both in $\overline{\lambda}$ and $\lambda$. Therefore
\begin{eqnarray}
   {\cal I}   & \sim & \int [dU] \exp[-S_{YM}]\sum_{i=o}^{\infty} \sum_{r+s=3i}(-1)^s { 1 \over (r!)^2
s!}(S_{Qn})^s \bigtriangleup \int [d \bar{\lambda} d\lambda]_c  
\nonumber\\
 & & ( S_{Qnc})^r (S_{Qcn})^r  \exp \left[{1 \over a} (\overline\lambda,( D + Q_H)\lambda)_c  \right] 
=  \sum_{i=o}^{\infty} \sum_{r+s=3i}{\cal I}_{r,s}. 
\end{eqnarray}
We remaind the reader that the symbol $\sim$ does not refer to any approximation, but it means that the
equality holds only under the Berezin integral.
The functions ${\cal I}_{r,s} $ can be evaluated by expanding the exponential wr to $Q_H$
\be
{\cal I}_{r,s} = \sum_{t=0}^{\infty} {\cal I}_{r,s,t}.
\ee
Since only even powers of $Q_H$ can contribute
\begin{eqnarray}
   {\cal I_{r,s,t}}   &=& \int [dU] \exp[-S_{YM}] (-1)^s { 1 \over (r!)^2 s! (2t)!}
(S_{Qn})^s \bigtriangleup \int [d \bar{\lambda} d\lambda]_c  
\nonumber\\
 & & ( S_{Qnc})^r (S_{Qcn})^r  \left( { 1\over a}(\overline\lambda, Q_H \lambda)_c \right)^{2t}
\exp \left[{1 \over a} (\overline\lambda, D \lambda)_c  \right].  
\end{eqnarray}

 Finally assuming the nucleon
composites as  integration variables we get the desired perturbative expansion of the partition function of QCD
in terms of the nucleon and chiral fields 
\begin{eqnarray}
 Z  & = &   \int [dA] \exp[-S_M - \mbox{Tr} \ln \left( C^{-1}(0) C(A) \right)] \int [d \bar{\psi} d\psi]_n
 \nonumber\\ 
 & & \int \left[{d\vec{\chi}\over\sqrt{2 \pi}}\right]_c  
   \left[{d\phi \over\sqrt{2 \pi}}\right]_c  \sum_{i=o}^{\infty} \sum_{r+s=3i} \;
 {\cal I}_{r,s }\exp[-S_N -S_{\chi}].
\end{eqnarray}
 The integrals over the quark and gluon fields are relegated in
the functions ${\cal I}_{r,s}$. Note that we do not need to treat the gluon field perturbatively: If and
where  this can possibly be done remains here an open question. 

At this point the role of $k_N^{-1},k_{\pi}^{-1}$ as expansion parameters should be clear. Since the integral
over the quarks  in the chiral lattice is proportional to $D^{-1}$, namely to $k_{\pi}^{-(2t+r)}$, and because
of the dependence of the transformation functions~\reff{h} on $k_N$, ${\cal I}_{r,s,t} \sim 
k_N^{-(r+s)/3} k_{\pi}^{-(2t+r)}$.

We conclude this Section by the evaluation of the correlation functions. To this aim we observe that the
Stratonovich-Hubbard transformation can be written in the form
\begin{eqnarray} 
& & \pi_{h_1}(c_1)...\pi_{h_s}(c_s) \exp[-S_C -S_{Qc}]  = \left(\det C(A) \right)^{-1} \int
\left[{d\vec{\chi}\over\sqrt{2 \pi}}\right]_c  
   \left[{d\phi\over\sqrt{2 \pi}}\right]_c \bigtriangleup \;
\nonumber\\
& & \exp [-S_{\chi}]  (a2a_t \rho^2)^{-s}{\partial \over \partial \chi_{h_1}(c_1)}...
{\partial \over \partial \chi_{h_s}(c_s)}
 \exp   \left [ {1 \over a}(\overline{\lambda},( D + Q_H )\lambda )_c \right]. 
\end{eqnarray}
Repeating the previous manipulations we then arrive at the explicit expression
\begin{eqnarray}
& &<\overline{\psi}(n_1)...\psi(n_r)\pi_{h_1}(c_1)...\pi_{h_s}(c_s)> = { 1\over Z}\int [dA]
\exp[-S_M - \mbox{Tr} \ln ( C^{-1}(0) C(A) )] 
\nonumber\\ 
 & & \;\;\;\;\;\;\;\;\;\;\;\;\;\;\; \int [d \bar{\psi} d\psi]_n \exp[-S_N]
\overline{\psi}(n_1)...\psi(n_r) \int \left[{d\vec{\chi}\over\sqrt{2 \pi}}\right]_c  
\left[{d\theta\over\sqrt{2 \pi}}\right]_c 
\nonumber\\
 & & \;\;\;\;\;\;\;\;\;\;\;\;\;\;\; \exp[-S_{\chi}] \bigtriangleup (a2a_t \rho^2)^{-s}
{\partial \over \partial \chi_{h_1}(c_1)}...{\partial \over \partial \chi_{h_s}(c_s)} 
\left(\bigtriangleup^{-1}{\cal I}\right).
 \end{eqnarray}

\section{Quark confinement}

Let us consider the quark propagator 
\begin{eqnarray}
 & &<\overline{\lambda}(x) \Gamma(x,y)\lambda(y)> = { 1\over Z} \int [dU] \int [dA]\, \exp[- S_{YM} - S_M] 
\nonumber\\
 & & \;\;\;\;\;\;\;\;\;\;\int [d\overline{\lambda} d\lambda]\; \overline{\lambda}(x) \Gamma(x,y)
\lambda(y)\exp[-S_N -S_{C}-S_Q],  
\end{eqnarray}
where $\Gamma(x,y)$ is an arbitrary string of gluons making the above integral gauge invariant. We can repeat 
the previous manipulations until we arrive at
\begin{eqnarray}
& &<\overline{\lambda}(x)\Gamma(x,y) \lambda(y)> = { 1\over Z}  \int [dA]  \exp[-S_M - \mbox{Tr} 
\ln ( C^{-1}(0) C(A) )]   \int [d \bar{\lambda} d\lambda]_n 
 \nonumber\\
  & & \;\;\exp[-S_N -S_{Qn}]\int
   \left[{d\vec{\chi}\over\sqrt{2 \pi}}\right]_c  
  \left[{d\theta\over\sqrt{2 \pi}}\right]_c  \exp[-S_{\chi}]\int [dU] \exp[-S_{YM}] \bigtriangleup
 \nonumber\\
 & &\;\;
 \int [d \bar{\lambda} d\lambda]_c \overline{\lambda}(x) \Gamma(x,y) \lambda(y) 
 \exp   \left[{1 \over a} (\overline{\lambda},(D +Q_H)\lambda)_c  - S_{Qnc} -S_{Qcn}\right].  
\end{eqnarray}
In the above formulation the quarks appear in interaction with the auxiliary fields and the nucleon fields,
and have an effective mass equal to $k_{\pi}( \rho^2 \overline{\phi} + \sqrt{\Omega}m) 
\sim \sqrt{\Omega} \rho k_{\pi}/ \sqrt{a2a_t}$. This mass seems to diverge in the continuum limit, but we will
see that the constant $k_{\pi}$ must be tuned with the lattice spacing in a way which will not be determined
in the present paper. What is relevant, however, as far as the perturbative expansion is concerned, is the
dependence on $\Omega$. Since the quark effective mass in the chirally broken vacuum grows like $\sqrt
{\Omega}$, the saddle point expansion results to be a strong coupling expansion for the quarks, so that if the
sites $x,y$ are $n$ lattice spacings apart from one another, the first nonvanishing contribution to the quark
propagator occurs at an order of the expansion not smaller than $n$. As a consequence the quarks are never
produced to any finite order, a fact first observed in~\cite{Cara1}. In this sense we can say that they are
confined. Since the quark effective mass to leading order in $\Omega$  does not depend on the breaking
parameter, {\it this confinement is a genuine consequence of the spontaneous chiral symmetry breaking}, in
whose absence there would be no saddle point expansion.

In conclusion in the present approach the quarks do not propagate at the perturbative level, so that they
might become alive only due to nonperturbative effects. The situation is reversed wr to the standard
perturbation theory where the quarks appear as physical particles whose confinement is intrinsically
nonperturbative.

Since the quarks have no poles whatsoever, it seems that we do not have to worry about the spurious ones. 
If this turns out to be confirmed by the study of the anomaly in the present approach, we can altogether
forget the Wilson term for the quarks. In any case we can safely assume the Wilson parameter $r_q$  of order
$1/ \sqrt{\Omega}$, and neglect the quark Wilson term to leading order in our expansion.

\section{The nucleon-pion interaction}

As an application we investigate the pion-nucleon interaction to lowest order. It occurs
in ${\cal I}_{1,2,0}$ , the lowest order term of ${\cal I}_{1,2}$. Since the integrand
will result to be independent on the gluon field, the corresponding integral will factor out and we will
therefore  omit it from the start  
\begin{eqnarray}
& &{\cal I}_{1,2,0} = { 1 \over 2}(S_{Qn})^2 \bigtriangleup \int [d \overline{\lambda}
d\lambda]_c  S_{Qnc} S_{Qcn}  \exp \left[{1 \over a} (\overline\lambda, D
\lambda)_c  \right] =  -{ 1\over 2} a^3 a_t (S_{Qn})^2
\nonumber\\
 & & \;\;\;\;\;\; \sum_{c,\epsilon, \eta}\left( Q_{c-e_{\epsilon t},c} D_c^{-1}Q_{c,c+e_{\eta t}}
\right)_{\overline{h}_1 h_1} \overline{\lambda}_{\overline{h}_1}(c-e_{\epsilon t}) \lambda_{h_1} (c+e_{\eta t}).
\end{eqnarray}
We look for the terms in ${\cal I}_{1,2,0}$ proportional to $\overline{\psi}, \psi$, namely containing three
$\lambda$ at one and the same site and also three $\overline{\lambda}$ at one and the same, possibly
different, site. They are obtained by retaining only the contributions with $\epsilon=-\eta$, and only the 
mass term from  $S_{Qn}$
\begin{eqnarray}
{\cal I}_{1,2,0} &=&  { 1\over 4} a^{10} a_t^3 m_Q^2 \sum_{c,\epsilon}
\left( Q_{c+e_{\epsilon t},c} D_c^{-1}Q_{c,c+e_{\epsilon t}} \right)_{\overline{h}_1 h_1}
\nonumber\\
 & &(\overline{\lambda}_{\overline{h}_1}\overline{\lambda}_{h_2}\overline{\lambda}_{h_3}
\lambda_{h_1} \lambda_{h_{_2}} \lambda_{h_3})_{c+e_{\epsilon t}}
\nonumber\\
 & =&{ 1\over 4} a^{10} a_t^3 M_Q^2 \sum_{c, \epsilon}
\left( Q_{c+e_{\epsilon t},c} D_c^{-1}Q_{c,c+e_{\epsilon t}} \right)_{\overline{h}_1 h_1}
\nonumber\\
  & & \overline{f_{\overline{I} \; \overline{h}_1 h_2 h_3 }} f_{I \; h_1 h_2
h_3}\overline{\psi}_{\overline{I}}(c+e_{\epsilon t}) \;\psi_I(c+e_{\epsilon t}). 
\end{eqnarray}
We need the transposed of the product $Q D^{-1} Q$. Assuming $r_q$ of the order $1 / \sqrt{\Omega}$ we can
neglect the Wilson term of the quarks so that
\begin{eqnarray}
  \left(Q_{c+e_t,c}   D_c^{-1}Q_{c,c+e_t}\right)^T &=& -{ 1\over 4 a_t^2} { 1\over D_c D_c^+}{\cal C} \tau_2
\gamma_t (v_t)^+(c) D_c
\nonumber\\
 & & \gamma_t v_t(c) {\cal C}^{-1} \tau_2 \otimes I_c,
\end{eqnarray}
where $I_c$ is the unit in color space.
Now by using the identities
\begin{eqnarray}
 & &\overline{f_{\overline{I} \; \overline{h}_1 h_2 h_3 }} f_{I \; h_1 h_2 h_3}(v_t^+ )_{ \overline{\tau}_1 } 
(v_t^+)_{ \tau_2}  (v_t^+)_{ \tau_3} v_{t \tau_1} v_{t \tau_2} v_{t \tau_3}
\nonumber\\
 & & = \overline{f_{\overline{I} \; \overline{h}_1 h_2 h_3 }} f_{I \; h_1 h_2 h_3}
 ( \hat{V}_t^+)_{ \overline{\tau}} \hat{V}_{t \tau}, \;\;\;\mbox{ no summation}
\end{eqnarray}
where
\be
\hat{V}_{t \tau}(c)= \exp (ie_{\tau} a_t A_t(c) ),
\ee
we get
\begin{eqnarray}
{\cal I}_{1,2,0} &=&  { 3\over 8} a^{10} a_t m_Q^2 \sum_{c, \epsilon}
{ 1\over D_c D_c^+} \left\{ {\cal C} \tau_2  \gamma_{\epsilon t} D_c
 \gamma_{\epsilon t} {\cal C}^{-1} \right\}_{h_1 \overline{h}_1}
\nonumber\\
 & & \overline{h_{\overline{I} \; \overline{h}_1 h_2 h_3}} \; 
h_{I \;  h_1 h_2 h_3} \left[\overline{\psi}(c+e_{\epsilon t}) \hat{V}^+(c)\right]_{\overline{I}}
\left[\hat{V}(c)\psi(c+e_{\epsilon t})\right]_I. 
\end{eqnarray}
The sum over colors has been performed. Using the explicit expressions of the functions $h$ 
\begin{eqnarray}
{\cal I}_{1,2,0} &=&    { 3\over 8} \; { 1 \over 24^2} a^4 a_t m_Q^2 \;  k_N^{-1} \sum_{c, \epsilon}
{ 1\over D_c D_c^+} 
\nonumber\\
 & &  \overline{\psi}(c+e_{\epsilon t}) \hat{V}^+(c)\left[ 3D_{1c} + i  D_{2c}\right] \hat{V}(c)
\psi(c+e_{\epsilon t}),  
\end{eqnarray}
where we have decomposed $D$ according to
\be
D= D_1 + i D_2.
\ee
Retaining the leading terms in $1/ \Omega$ in the matrix $D$ and summing over $\epsilon$ we get the final
result  
\begin{eqnarray}
{\cal I}_{1,2,0} &=&  i g_{N-\pi} a^3 2a_t \sum_c
\overline{\psi}(c+e_t) \hat{V}^+(c)  \gamma_5 \tau \cdot \chi(c) \hat{V}(c)\psi(c+e_t)
\nonumber\\
& & + a^3 2a_t \sum_c \delta m_N \overline{\psi}(c+e_t) \psi(c+e_t), 
\end{eqnarray}
where 
\be
g_{N-\pi}= { 3\over 8} \; { 1 \over 24^2} { 2a_t \over a}  { a^2 m_Q^2 \over \Omega k_N k_{\pi}} 
\ee
is the nucleon-pion coupling constant and
\be
\delta m_N={ 9 \over 8} \; { 1 \over 24^2} \sqrt{{ 2a_t \over a}}{ a  m_Q  \over \sqrt{\Omega} \rho k_N
k_{\pi}}  m_Q
\ee
is a renormalization of the nucleon mass. In the continuum limit the shift in the arguments of the fields can
be ignored, and the link variables $\hat{V}$ disappear.

The interpretation of $g_{N-\pi}$ as the pion-nucleon coupling constant follows by the evaluation of the
3-point correlation function 
\begin{eqnarray}
& &<\overline{\psi}(n_1)\psi(n_2)\pi_{h}(c)> = { 1\over Z}\int [dA]
\exp[-S_M - \mbox{Tr} \ln ( C^{-1}(0) C(A) )] 
\nonumber\\ 
 & & \;\;\;\;\;\;\;\;\;\;\;\;\;\;\; \int [d \bar{\psi} d\psi]_n \exp[-S_N]
\overline{\psi}(n_1)\psi(n_2) \int \left[{d\vec{\chi}\over\sqrt{2 \pi}}\right]_c  
\left[{d\theta\over\sqrt{2 \pi}}\right]_c 
\nonumber\\
 & & \;\;\;\;\;\;\;\;\;\;\;\;\;\;\; \exp[-S_{\chi}] \bigtriangleup (a2a_t \rho^2)^{-1}
{\partial \over \partial \chi_{h}(c)} \left(\bigtriangleup^{-1}{\cal I}_{1,2,0} \right).
 \end{eqnarray}
Since the term with the derivative of ${\cal I}_{1,2,0}$ does not contribute in the continuum limit, we
have
\begin{eqnarray}
& &<\overline{\psi}(n_1)\psi(n_2)\pi_{h}(c)> = { 1\over Z}\int [dA]
\exp[-S_M - \mbox{Tr} \ln ( C^{-1}(0) C(A) )] \int [d \bar{\psi} d\psi]_n 
\nonumber\\ 
 & &  
\overline{\psi}(n_1)\psi(n_2) \exp[-S_N]\int \left[{d\vec{\chi}\over\sqrt{2 \pi}}\right]_c  
\left[{d\theta\over\sqrt{2 \pi}}\right]_c \chi_{h}(c) {\cal I}_{1,2,0} \exp[-S_{\chi}] .
 \end{eqnarray}

\section{Conclusions}

We have derived a perturbative expansion in QCD in terms of the quark composites with the quantum numbers
of the nucleons and the auxiliary fields which replace the chiral composites. The coefficients appearing
in the perturbative series are defined in terms of integrals over the quark and gauge fields.

The expansion parameters are the inverse of $\Omega$, the number of quark components, the inverse of the
constants $k_N, k_{\pi}$, entering the definition of the composites and the quark mass $m_Q$. The nature of the
expansion is however at present not fully understood: $k_N$ is indeed accompanied by large numerical factors
related to the number of quark components, but  we do not know how the powers of these factors will be related
to the powers of $k_N$ in higher order terms. We do not see at present any other way of clarifying this issue
than by studying higher order terms.

To leading order the nucleon-pion interaction does not involve the gluon field, so that it is not affected by
the gluon dynamics, but it is  solely determined by the hadronic quantum numbers. We regard this as the result
of the choice of the "right" variables, aimed at suppressing the degrees of freedom irrelevant in the regime
of interest.

One of the problems where the present formulation of QCD seems particularly promising is the phase transition
at high barion density~\cite{QCD}. The reason is that in the usual way of introducing the chemical potential
this is coupled to the quarks to allow an analytical integration to be performed on the latters. But the price
to be paid is an interplay between gluon field and chemical potential which would not be there if we coupled
the chemical potential to the nucleons, what we can easily do in the present formalism.

\end{document}